\documentclass[traditabstract]{aa}
\usepackage{natbib}
\usepackage{epsfig}

\def\gtsima{$\; \buildrel > \over \sim \;$}
\def\ltsima{$\; \buildrel < \over \sim \;$}
\def\gtrsim{\lower.5ex\hbox{\gtsima}}
\def\lesssim{\lower.5ex\hbox{\ltsima}}

\begin{document}
%

\title{Building gas rings and rejuvenating \\S0 galaxies through minor mergers}
\author{Michela Mapelli\inst{1}, Roberto Rampazzo\inst{1}, Antonietta Marino\inst{2}} 
\institute{INAF-Osservatorio Astronomico di Padova, Vicolo dell'Osservatorio 5, I--35122, Padova, Italy\\ \email{michela.mapelli@oapd.inaf.it}
\and
Dipartimento di Fisica e Astronomia Galileo Galilei, University of Padova, Vicolo dell'Osservatorio 3,  I--35122, Padova, Italy}

 \titlerunning{Building gas rings in S0 galaxies}
 
\authorrunning{Mapelli et al.}
 
\abstract{
We investigate the effects of minor mergers between an S0 galaxy and a gas-rich satellite 
galaxy, by means of N-body/smoothed particle hydrodynamics (SPH) simulations. 
The satellite galaxy is initially on a nearly parabolic orbit and undergoes several 
periapsis passages before being completely stripped. In most simulations, 
a portion of the stripped gas forms a warm dense gas ring in the S0 galaxy, with 
a radius of $\sim{}6-13$ kpc and a mass of $\sim{}10^7$ M$_\odot$. 
The ring is generally short-lived ($\lesssim{}3$ Gyr) if it forms from prograde 
encounters, while it can live for more than 6 Gyr if it is born from 
counter-rotating or non-coplanar interactions. The gas ring keeps 
memory of the initial orbit of the satellite galaxy: it is corotating (counter-rotating) with the stars of the disc of the S0 galaxy, if it originates from 
prograde (retrograde) satellite orbits. Furthermore, 
the ring is coplanar with the disc of the S0 galaxy only if the satellite's orbit 
was coplanar, while it lies on a plane that is inclined with respect to 
the disc of the S0 galaxy by the same inclination angle as the orbital 
plane of the satellite galaxy. The fact that we form polar rings as 
long-lived and as massive as co-planar rings suggests that rings 
can form in S0 galaxies even without strong bar resonances. 
Star formation up to 0.01 M$_\odot{}$ yr$^{-1}$ occurs for $>6$ Gyr 
in the central parts of the S0 galaxy as a consequence of the interaction. 
We discuss the implications of our simulations for the rejuvenation of 
S0 galaxies in the local Universe.
}

\maketitle

%

\section{Introduction}


A large fraction of S0 galaxies ($\sim{}20-25$\%{}, \citealt{Laurikainen2013}) show signatures of rings (i.e. closed structures composed of stars, gas and dust) and pseudo-rings (incomplete versions of rings, sometimes formed by arm-like structures, \citealt{Comeron2014}). Far ultraviolet (FUV) observations frequently evidence rings and pseudo-rings of gas and/or young stars  in early-type galaxies (ETGs), and especially in S0 galaxies (\citealt{Thilker2007,Salim2010,Salim2012}).
\cite{Salim2010} analyze 22 ETGs with an extended 
ultraviolet (UV) morphology in {\it GALEX} data, and find
rings in 15 of them. \citet{Salim2012} investigate a sample of ETGs at $z<0.1$, 
27 of which have been observed with the {\it Hubble Space Telescope} ({\it HST}) in FUV. These ETGs are UV structured, showing ring 
and/or arm-like structures, while in the optical they have an old stellar disc, i.e. they are S0 galaxies. The UV emission is consistent with a low-level galaxy-scale star formation (SF) of $\sim{}0.5$ M$_{\odot}$ yr$^{-1}$.  

Rings in S0 galaxies are found at different  distances from the galaxy nucleus, often connected to the presence 
of bar/s \citep[see e.g.][]{Aguerri2009,Laurikainen2013}. Such kind of rings are often classified, depending on their distance from the nucleus, as (i) nuclear rings (surrounding the nucleus of the galaxy), 
 (ii) inner rings (generally surrounding the bar, if present), and (iii) outer rings (more external than the  former ones). 

The origin of the rings is still debated. The fact that many rings are associated with  bars 
(or other non-axisymmetric structures) has led to the idea that the formation of rings is connected with 
resonances induced by the rotation of the bar (i.e. either an inner Linblad's resonance, or an outer
 Linblad's resonance, or a ultraharmonic resonance, \citealt{Binney1987, butacombes1996}). 
 However, a resonance cannot be invoked for the outer polar rings (e.g. \citealt{Marino2009}), as 
 well as for the most asymmetric and off-centre rings (e.g. \citealt{Thilker2010}). In addition, bars 
 should be more common than observed in S0 galaxies ($\sim{}30$\%, \citealt{Aguerri2009}) to 
 explain the large fraction of ringed S0 galaxies found in \cite{Salim2010}.

Alternative scenarios to a bar-driven resonance include gas-rich minor or major mergers, 
and accretion from filaments/clouds of intergalactic medium.
 Mergers and gas accretion from filaments are the only models able to explain the most massive polar rings (\citealt{Bekki1998,Iodice2002, Iodice2006, Moiseev2011, Spavone2010, Spavone2011,Spavone2013}). Evidences of gas accretion either from filaments or from interactions with gas-rich satellite galaxies have been found in a large number of galaxies (e.g. \citealt{deBlok2014, serraoosterloo2010}; see \citealt{Sancisi2008} for a review). Similarly, the analysis of a large sample of disc galaxies in the Sloan Digital Sky Survey reported by \citet{Kaviraj2014a} suggests that minor mergers played a major role in driving SF in galaxy discs (see also \citealt{Kaviraj2010,Kaviraj2012,Kaviraj2013,Kaviraj2014b}). Numerical simulations also indicate the importance of minor mergers for the accretion of fresh gas in ETGs (e.g. \citealt{Kaviraj2009, Peirani2010}). Finally, \cite{Marino2011} investigate five S0 galaxies with outer rings in UV (NGC~1533, NGC 2962, NGC 2974, NGC 4245, and NGC 5636). While all the considered galaxies are barred and the rings might be related to bar-induced resonances, there are several indications that NGC~1533, NGC~2962 and NGC~2974 have accreted gas from satellite galaxies. Thus, \cite{Marino2011} hypothesize that the outer rings in their sample are an effect of (bar-driven) secular evolution, but at the same time this process was aided by fresh gas accreted from gas-rich satellites.  Using N-body/smoothed-particle hydrodynamics (SPH) simulations with chemo-photometric implementation, \citet{Mazzei2014} showed that the pseudo-ring in NGC~1533 and its star forming properties may be a transient phase during a major merger episode involving two halos with  mass ratio  2:1.



In this paper, we study the formation of gas rings in S0 galaxies, by means of N-body/SPH simulations of minor mergers with small gas-rich companions. 
In Section~2, we describe the adopted numerical techniques. In Section~3 we present our results, with particular attention for the formation of co-planar and polar gas rings. In Section~4 we discuss the main implications of our results, while in Section~5 we summarize our conclusions.

\begin{table}
\begin{center}
\caption{Initial conditions of the $N-$body simulations: masses and scale lengths.}
 \leavevmode
\begin{tabular}[!h]{lll}
\hline
Model galaxy properties & Primary & Secondary \\
\hline
$M_{\rm DM}$ ($10^{11}$ M$_\odot{}$)          & 7.0  & 0.3\\
$M_\ast{}$ ($10^{10}$ M$_\odot{}$)      & 7.0   & 0.2\\
$f_{\rm b/d}$                               &  0.25 & 0.25 \\
$M_{\rm G}$ ($10^{8}$ M$_\odot{}$)    & 0 &  1.38 \\
$R_{\rm s}$ (kpc) & 6.0 & 3.0\\
$R_{\rm d}$ (kpc) & 3.7 & 3.0 \\
$h_{\rm d}$ (kpc) & 0.37 & 0.30 \\
$r_{\rm B}$ (kpc) & 0.6 & 0.6 \\
\noalign{\vspace{0.1cm}}
\hline
\end{tabular}
\begin{flushleft}
\footnotesize{$M_{\rm DM}$ is the DM mass; $M_\ast{}$ is the total stellar mass of the galaxy (including both bulge and disc); $f_{\rm b/d}$  is the bulge-to-disc mass ratio; $M_{\rm G}$ is the total gas mass. The primary has no gas, while the gas of the secondary is distributed according to an exponential disc (\citealt{Hernquist1993}), with the same parameters (scale length and height) as the stellar disc. $R_{\rm s}$ is the NFW scale radius $R_{\rm s}\equiv{}R_{200}/c$, where $R_{200}$ is the virial radius of the halo \citep{Navarro1996} and $c$ the concentration (here we assume $c=12$ for both galaxies). $R_{\rm d}$ and $h_{\rm d}$ are the disc scale length and height, respectively (\citealt{Hernquist1993}), while $r_{\rm B}$ is the bulge scale length.}
\end{flushleft}
\end{center}
\end{table}

\begin{table*}
\begin{center}
\caption{Initial conditions of the $N-$body simulations: orbital properties.}
 \leavevmode
\begin{tabular}[!h]{cccccccc}
\hline
  Run &  $b$  & $v_{\rm rel}$ & $\theta{}$, $\phi{}$, $\psi{}$ & $E_{\rm s}$  & $L_{\rm s}$ & $e$  & Orbit spin \\
  &  (kpc) & (km s$^{-1}$) & (rad)  & (10$^4$ km$^2$ s$^{-2}$) & (10$^3$ km s$^{-1}$ kpc) & & \\
\hline
A        & 10.0   & 200 & $-\pi{}/2$, 0, 0      & 0.38  & 2.0 &  1.003 &  retrograde \\
B        & 10.0   & 200 & $\pi{}/2$, 0, 0       & 0.38  & 2.0 &  1.003 &  prograde \\
C        & 10.0   & 200 & $-\pi{}/4$, 0, 0      & 0.38  & 2.0 &  1.003 & -- \\
D        & 10.0   & 200 & 0, 0, 0               & 0.38  & 2.0 &  1.003 & -- \\
E        & 30.0   & 200 & $-\pi{}/2$, 0, 0      & 0.38  & 6.0 &  1.03  & retrograde\\
F        & 30.0   & 200 & $\pi{}/2$, 0, 0       & 0.38  & 6.0 &  1.03  & prograde \\
\noalign{\vspace{0.1cm}}
\hline
\end{tabular}
\begin{flushleft}
\footnotesize{$b$ and  $v_{\rm rel}$ are the impact parameter and the relative velocity between the centres of mass (CMs) of the two galaxies at the initial distance, respectively. For the definition of $\theta{}$, $\phi{}$, $\psi{}$, see figure~1 of \cite{Hut1983}. In particular, $\theta{}$ is the angle between the relative velocity vector ${\bf v}_{\rm rel}$ and the symmetry axis of the primary disc, $\phi{}$ describes the orientation of  ${\bf v}_{\rm rel}$ projected in the plane of the primary disc and $\psi{}$ describes the orientation of the initial distance vector ${\bf D}$ (between the CMs of the two galaxies) in the plane perpendicular to ${\bf v}_{\rm rel}$.\\
 $E_{\rm s}$ is the  specific orbital energy, i.e. the total energy divided by the reduced mass $\mu{}=m_1\,{}m_2/(m_1+m_2)$, where $m_1$ and $m_2$ are the mass of the primary and of the secondary galaxy, respectively. $E_{\rm s}\equiv{}-G\,{}M/D+\,{}v_{\rm rel}^2/2$, where $M=m_1+m_2$ is the total mass of the two galaxies, 
$G$ is the gravitational constant and $D$ the initial distance between the CMs. \\$L_{\rm s}$ is the modulus of the specific orbital angular momentum, i.e. the angular momentum divided by the reduced mass. \\$e$ is the eccentricity ($e=[1+2\,{}E_{\rm s}\,{}L_{\rm s}^2/(G\,{}M)^2]^{1/2}$). \\A orbit is classified as prograde/retrograde depending on the alignment/counter-alignment of the orbital angular momentum of the secondary galaxy with respect to the spin of the primary galaxy.}
\end{flushleft}
\end{center}
\end{table*}

\section{Method: $N-$body simulations}\label{sec:method} 
We simulate interactions between a primary S0 galaxy and a secondary gas-rich galaxy. The initial conditions for both the primary galaxy and the secondary galaxy in the $N-$body model are generated by using an upgraded version of the code described in Widrow, Pym \&{} Dubinski (2008; see also \citealt{Kuijken95} and \citealt{Widrow2005}). The code generates self-consistent disc-bulge-halo galaxy models, derived from explicit distribution functions for each component, which are very close to equilibrium. In particular, the halo is modelled with a Navarro, Frenk \&{} White (1996, NFW) profile. We use an exponential disc model \citep{Hernquist1993}, while the bulge is spherical and comes from a generalization of the Sersic law (\citealt{Prugniel1997, Widrow2008}).
 
Both the primary and the secondary galaxy have a stellar bulge and a stellar disc.
 The giant S0 galaxy has no gas, whereas the secondary galaxy has an initial gas mass of $1.38\times{}10^{8}$ M$_\odot{}$, distributed according to an exponential disc. Therefore, the initial configuration of the secondary galaxy is consistent with a low-mass gas-rich disc galaxy. The assumption that the primary has no gas is quite unrealistic, but allows us to easily distinguish between the triggering of SF in the pre-existing gaseous disc of the S0 galaxy and the effect of fresh gas accretion from the companion galaxy.

The total mass of the secondary is $\sim{}1/20$ of the mass of the primary, classifying the outcome of the interaction as a minor merger. 
The masses of the various components and the scale lengths of the simulated galaxies are listed in Table~1. 
Table~2 shows the orbital properties (impact parameter, relative velocity, orientation angles and total energy) of the six runs. 
The adopted orbits are nearly parabolic (Table~2), in agreement with predictions from cosmological simulations \citep{Khochfar2006}. The main implication of such  nearly parabolic orbits is that the stripping of the satellite galaxy proceeds slowly: the merger is not yet complete in most simulations at $t=11$ Gyr since the first periapsis passage.

In all the simulations, the particle mass in the primary galaxy is $2.5\times{}10^5$ M$_\odot{}$ and $5\times{}10^4$ M$_\odot{}$ for dark matter (DM) and stars, respectively. The particle mass in the secondary galaxy is  $2.5\times{}10^4$ M$_\odot{}$ for DM and $5\times{}10^3$ M$_\odot{}$ for both stars and gas. 
The softening length is $\epsilon{}=100$ pc.

We simulate the evolution of the models with the $N-$body/SPH tree code gasoline \citep{Wadsley2004}. Radiative cooling, SF and supernova blastwave feedback are enabled, as described in Stinson et al. (2006, 2009, see also \citealt{Katz1992}). The adopted parameters for SF and feedback  are the same as used in recent cosmological simulations capable of forming realistic galaxies in a wide range of masses (e.g., \citealt{Governato2010,Guedes2011}), and in recent simulations of galaxy-galaxy collisions (\citealt{Mapelli2012a,Mapelli2012b,Mapelli2013a,Mapelli2013b}).

In runs~A and B the orbit of the satellite galaxy is co-planar with the disc of the S0 galaxy. The only difference between run~A and B is that the orbit of the satellite galaxy is retrograde  and prograde (i.e. the orbital angular momentum of the satellite is counter-aligned and aligned with respect to the spin of the primary galaxy) in run~A and B, respectively (Table~2).
 Run A, C and D differ for the inclination angle between the relative velocity vector ${\bf v}_{\rm rel}$ and the symmetry axis of the primary disc ($\theta{}$, see Table 2): $\theta{}=-\pi{}/2, -\pi{}/4$ and 0 for run A, C and D, respectively. This means that the orbit of the satellite galaxy is inclined by 45$^{\circ{}}$ and 90$^{\circ{}}$ with respect to the disc of the primary galaxy in run C and D, respectively.

In runs E (retrograde) and F (prograde), the impact parameter (30 kpc) is a factor of 3 larger than in run A. 
In all simulations, the primary galaxy develops a strong bar $\sim{}200$ Myr after the beginning of the simulation. Our galaxy models are marginally stable against the formation of a bar: evolving the galaxy in isolation, the bar does not appear. The perturbation induced by the interaction with the intruder satellite galaxy plays an important role in the formation of the bar.

\section{Results}
\begin{figure}
\center{{
\epsfig{figure=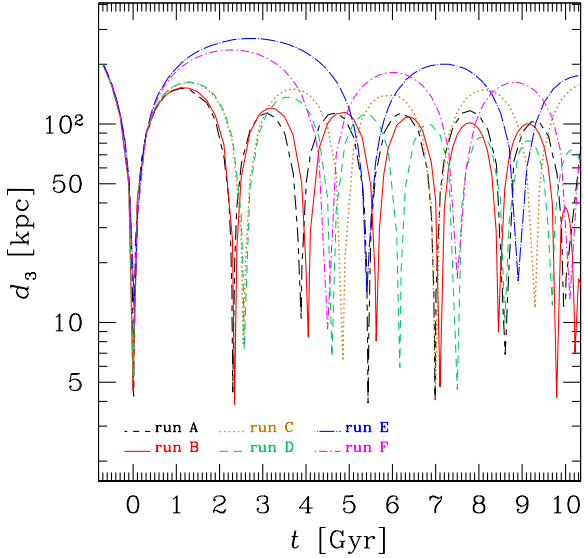,width=8.0cm} 
}}
\caption{\label{fig:fig1}
Radial distance of the satellite galaxy from the centre of the primary galaxy for all simulations, as a function of time. Time $t=0$ corresponds to the first periapsis passage. Long-short dashed black line: run~A; solid red line: run~B; dotted ochre line: run~C; dashed green line: run~D; long-dash-dotted blue line: run~E; short-dash-dotted magenta line: run~F.
}
\end{figure}
\begin{figure}
\center{{
\epsfig{figure=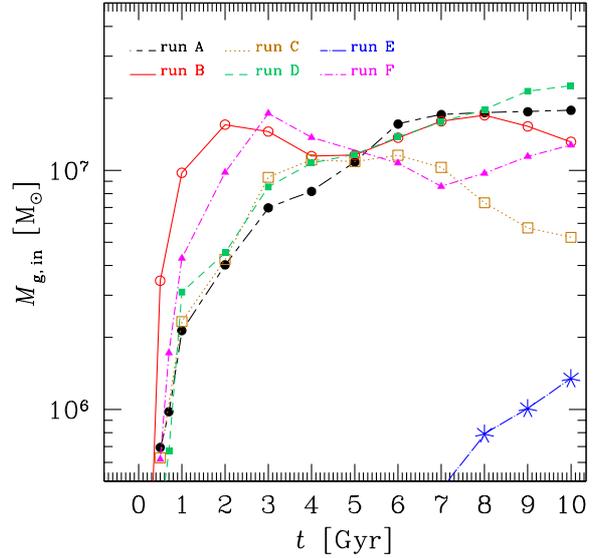,width=8.0cm} 
}}
\caption{\label{fig:fig2}
Mass of the gas that lies in the inner 15 kpc of the S0 galaxy ($M_{\rm g,\,{}in}$) for all simulations, as a function of time. Time $t=0$ corresponds to the first periapsis passage. Long-short dashed black line and filled black circles: run~A; solid red line and open red circles: run~B; dotted ochre line and open ochre squares: run~C; dashed green  line and filled green squares: run~D; long-dash-dotted blue line and blue asterisks: run~E;  short-dash-dotted magenta line and filled magenta triangles: run~F.
}
\end{figure}
\begin{figure}
\center{{
\epsfig{figure=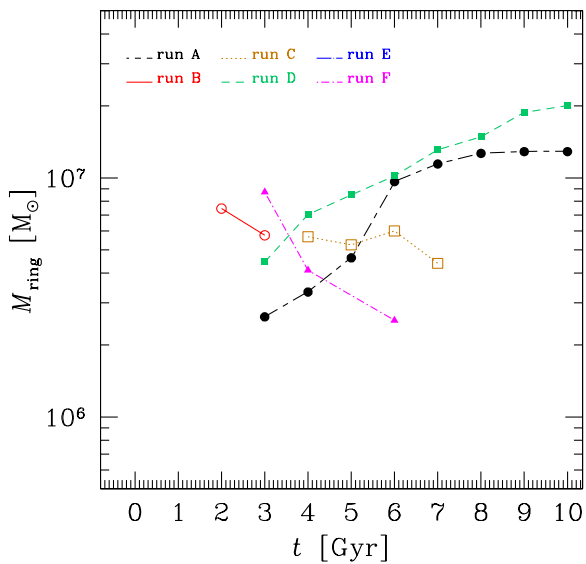,width=8.0cm} 
}}
\caption{\label{fig:fig3}
Mass of the gas ring ($M_{\rm ring}$) for all simulations, as a function of time. Time $t=0$ corresponds to the first periapsis passage. Long-short dashed black line and filled black circles: run~A; solid red line and open red circles: run~B; dotted ochre line and open ochre squares: run~C; dashed green  line and filled green squares: run~D; short-dash-dotted magenta line and filled magenta triangles: run~F.
}
\end{figure}
Fig.~\ref{fig:fig1} shows the radial distance of the satellite galaxy from the centre of the primary galaxy for all simulations, as a function of time (in this Figure and in the following, time $t=0$ corresponds to the first periapsis passage). From this figure, we can derive information about the periapsis passages. The impact parameter is extremely important to determine the amount of tidal disruption of the satellite at the first periapsis passage, and the subsequent orbital decay. In fact, the orbital decay is slower in the two runs with the largest impact parameter: in 10 Gyr, the satellite undergoes only 2 and 3 periapsis passages in runs~E and F, respectively. In the other runs, the satellite undergoes between 4 and 8 periapsis passages in 10 Gyr. Generally, the orbital decay is faster for prograde satellites (runs~B and F) than for their retrograde analogs (runs~A and E): tidal disruption occurs faster in prograde encounters, because of the lower relative velocity between the stars of the primary galaxy and the gas and stars of the secondary galaxy \citep{Toomre1972}. 
\subsection{The gas rings}
At each periapsis passage, gas and stars are stripped away from the satellite galaxy, and remain bound to the primary galaxy. In some runs, a portion of the stripped gas settles into a warm gas ring ($T\sim{}10^4$ K), with an average radius of $\sim{}6-13$ kpc. Figure~\ref{fig:fig2}  shows the total mass of stripped gas that resides in the inner 15 kpc of the primary galaxy ($M_{\rm g,\,{}in}$), as a function of time. By definition, $M_{\rm g,\,{}in}$ includes all the gas that lies in the inner 15 kpc of the S0 galaxy, including diffuse hot gas, the gas that is funnelled in the nucleus of the galaxy, and the gas that is confined into the ring. $M_{\rm g,\,{}in}$ becomes non-negligible ($\gtrsim{}2\times{}10^6$ M$_\odot$) soon after the first periapsis passage ($t\sim{}0.5-1.0$ Gyr) in all simulations apart from run~E. In run~E (retrograde and with a large impact parameter), gas stripping starts only in the last Gyrs.  In all simulations apart from run~E, $M_{\rm g,\,{}in}$ reaches a maximum value of $1-2\times{}10^{7}$ $M_\odot{}$ (i.e. $\sim{}1/10$ of the total gas mass of the satellite galaxy). From Figure~\ref{fig:fig2}, we also notice that the peak value of $M_{\rm g,\,{}in}$ is reached earlier in the prograde runs (run~B and F) than in the retrograde runs (runs~A and E) and in the non-coplanar runs (runs~C and D). This confirms that gas stripping is more efficient in the prograde runs. The fact that retrograde runs and  non-coplanar runs behave in a similar way is remarkable.

Figure~\ref{fig:fig3} shows the total mass of stripped gas that is confined in a gas ring ($M_{\rm ring}$), as a function of time. During the ring's life, the value of $M_{\rm ring}$ spans from 20\%  to 90\%  of $M_{\rm g,\,{}in}$, depending on the run. From Figure~\ref{fig:fig3}, it is apparent that the formation of a ring, its survival, and the amount of mass that is locked into the ring strongly depend on the characteristics of each run. In some runs, the ring never forms (run~E), or is relatively short-lived ($1-3$ Gyr, runs~B, C and F), while in other runs the ring grows steadily and is long-lived ($>7$ Gyr, runs~A and D).
\begin{figure*}
\center{{
\epsfig{figure=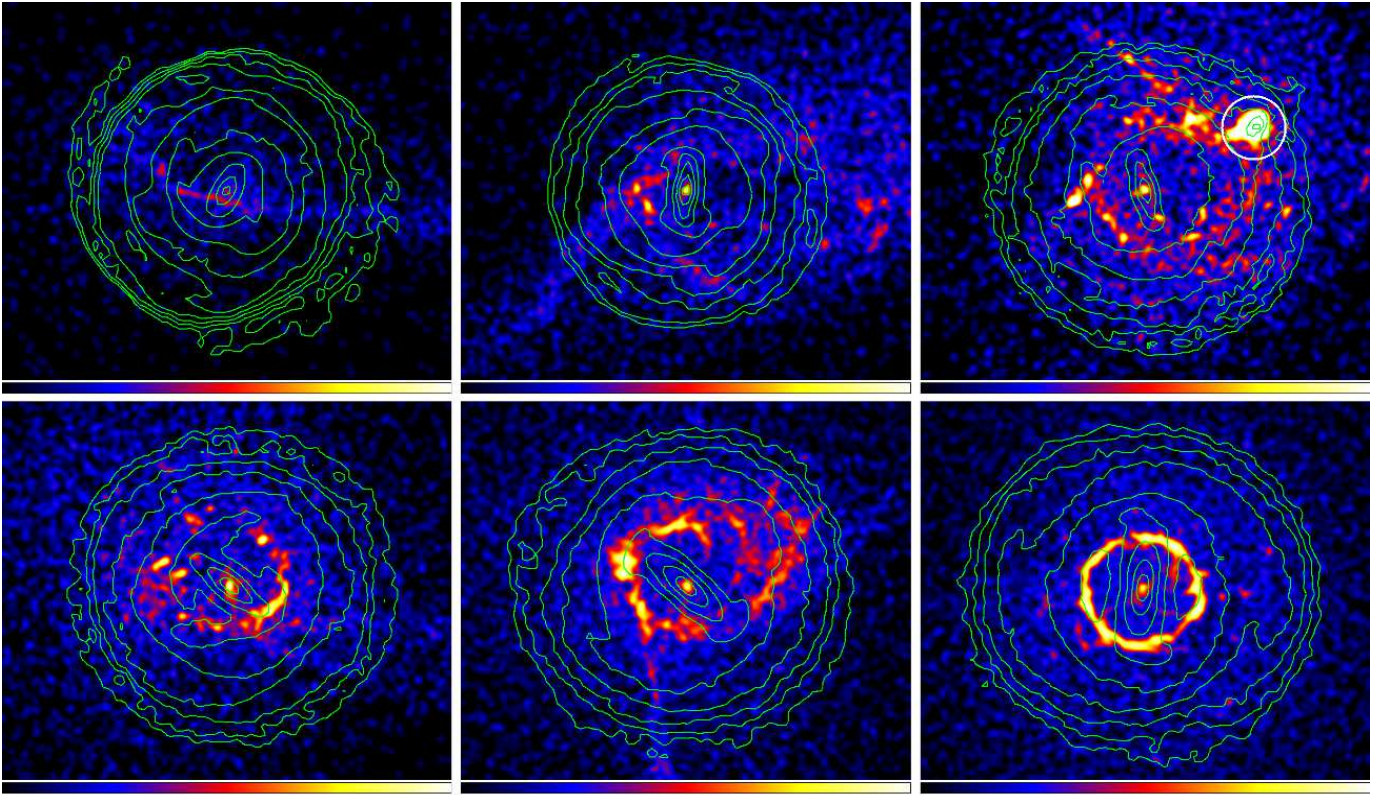,width=18.0cm} 
}}
\caption{\label{fig:fig4}
Projected density of gas (logarithmic colour-coded map) and stars (isocontours) of the primary galaxy in run A. From top left to bottom right: 1, 2, 2.35, 3, 5, and 7 Gyr after the first periapsis passage. Each panel measures $79\times{}67$ kpc. In the top right-hand panel, the white circle marks the position of the nucleus of the satellite galaxy, $\sim{}50$ Myr after the second periapsis passage. 
}
\end{figure*}

In run~A (Fig.~\ref{fig:fig4})  a co-planar ring of gas forms in the disc of the S0 galaxy at $t\sim{}3$ Gyr, i.e. approximately after the second periapsis passage. The ring  has a radius of $\sim{}8-10$ kpc, which means that it forms just at the end of the stellar bar, but is counter-rotating with respect to the disc, because it keeps memory of the retrograde orbit of the satellite. The ring in run~A is extremely long-lived: it survives for the entire simulation (Figure~\ref{fig:fig3}). In contrast, a gas ring forms earlier (at $t\sim{}2$ Gyr) in run~B (the prograde analogous of run~A), is co-rotating with the stellar disc, and survives only for $\sim{}1$ Gyr (Figure~\ref{fig:fig3}). The main reason for the early disappearance of the ring in run~B is that most of the gas is channelled inwards by the bar, and ends up in the nucleus.  



\begin{figure}
\center{{
\epsfig{figure=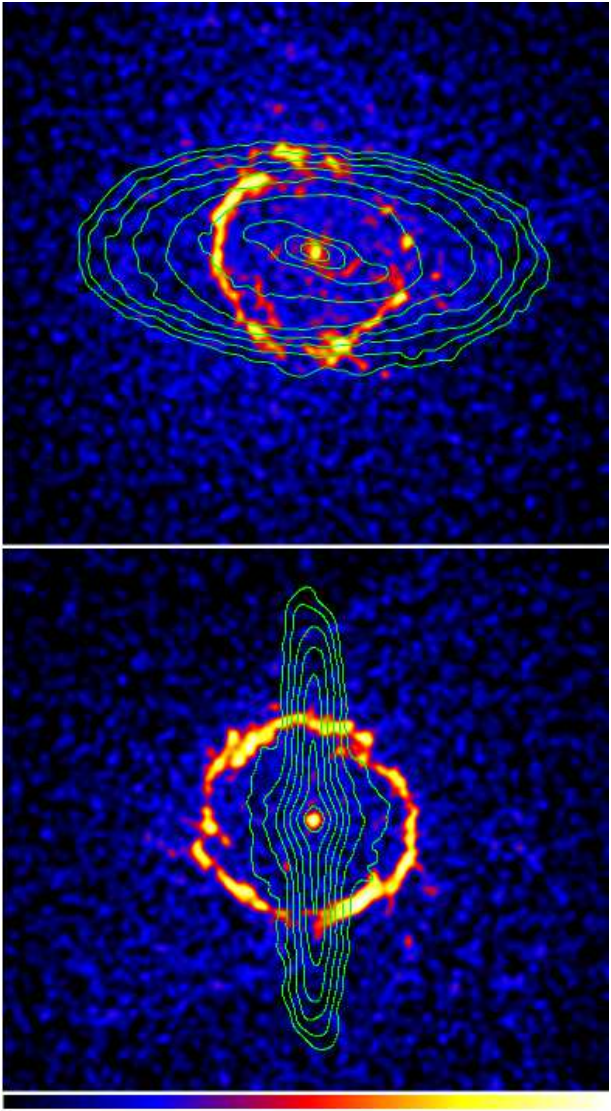,width=8.0cm} 
}}
\caption{\label{fig:fig5}
Projected density of gas (logarithmic colour-coded map) and stars (isocontours) of the primary galaxy 4 Gyr after the first periapsis passage. Top panel: run C. Bottom panel: run D. Each panel measures $79\times{}67$ kpc.
}
\end{figure}

The fact that the radius of the ring coincides almost with the size of the bar in runs~A and B might lead to the conclusion that the ring is the consequence of a bar-driven resonance. On the other hand, rings form even in runs~C and D (Figure~\ref{fig:fig5}), in which the orbit of the satellite is inclined by 45 and 90$^\circ$ with respect to the disc of the primary galaxy, respectively. Actually, it seems that the ring preserves the inclination of the orbit of the satellite with respect to the primary galaxy, as the gas ring of run~C and D forms with an inclination of  $\sim{}45$ and $\sim{}90^\circ$ with respect to the disc of the primary galaxy. In particular, the gas ring of run~D is a true polar ring, and cannot be produced by a bar-driven resonance. The ring of  run~D grows steadily in mass and survives till the end of the simulation. The radius of the ring in both run~C and D is $\sim{}12-13$ kpc, slightly larger than in the other runs. 

 The fact that the ring has approximately the same orientation as the initial satellite orbit is consistent with our expectations: no dynamical process (e.g. bar-driven resonance, differential precession) is sufficiently efficient to perturb the initial orbit of gas and stellar streams within a few Gyr, if the satellite's orbit has a large inclination. Even shocks and virial heating (e.g. \citealt{Keres2005}) are partially inefficient for our systems, given the lack of gas in the halo of the S0 galaxy. Finally, \citet{Theis2006} find that spiral arm-like perturbations might develop in a  polar ring surrounding a S0 galaxy, but these perturbations are confined in the initial plane of the polar ring.  We warn that our results are valid only for spherical halos. A triaxial halo is expected to exert torques, and to induce warps and tilts in the stellar disc (e.g. \citealt{Sparke1984,David1984,Steiman1984,Nelson1995,Dubinski2009}). Such effects might perturb a polar gas ring significantly. 

Furthermore, our findings are consistent with previous studies which investigate the formation of polar rings from the disruption of gas-rich satellite galaxies (e.g. \citealt{Richter1994,Mazzei2014}). In particular, there is consensus that the initial plane of the ring is the orbital plane of the companion if this is promptly destroyed (e.g. \citealt{Katz1992, Thakar1996}). The initial ring plane might then evolve to a preferred plane (either the equatorial or a polar plane) on a timescale longer than the orbital or azimuthal smearing time (i.e. several Gyr, e.g. \citealt{Rix1991,Cox1996}, see also  \citealt{Struck1999}, and references therein). Based on a sample of 78 polar ring galaxy candidates, \cite{Smirnova2013} recently proposed that the DM halo and the stellar bar are crucial to stabilize outer and inner polar rings, respectively. This is fairly consistent with our results, which suggest that the formation of the ring is not due to bar-driven resonance, but the bar and the DM halo contribute to stabilize the ring, and make it long-lived.



In both runs E (retrograde) and F (prograde) the orbit of the satellite is co-planar with the disc of the primary galaxy. The main difference with runs~A and B is that the periapsis passage is 3 times larger in runs~E and F. 
Only in run~E does no significant ring form over the  duration of the simulation ($t\sim{}10$ Gyr), because the periapsis distance of the satellite is too large and the relative velocity too high to produce significant stripping. 

In run~F, which is the prograde analogous to run~E,  a ring forms already at $t\sim{}3$ Gyr since the first periapsis passage (Figure~\ref{fig:fig3}). This confirms that gas stripping is much more efficient if the satellite orbit is prograde. Despite the larger periapsis with respect to the other runs, the ring of  run~F forms again at the end of the stellar bar, with a radius of $\sim{}10-12$ kpc. The ring of run~F disappears quite fast ($t\sim{}6$ Gyr). This result is analogous to what we already discussed for run~B: in the prograde runs the ring appears earlier than in the retrograde ones, but disappears earlier.

\subsection{The long-lived gaseous and stellar halo}\label{sec:halo}
 A considerable fraction of the initial stellar and gaseous mass of the satellite galaxy is scattered into the halo of the S0 galaxy, as a consequence of the encounter.  Fig.~\ref{fig:figextra} shows the total mass of stars  ($M_{\rm \ast,\,{}out}$) and gas ($M_{\rm g,\,{}out}$) that was initially bound to the satellite galaxy and that, at time $t$, lies at a distance $>50$ kpc both from the centre of the S0 galaxy and from the centre of the satellite galaxy. In our definition of $M_{\rm \ast,\,{}out}$ and $M_{\rm g,\,{}out}$, we exclude all stars and gaseous particles that lie at a distance $<50$ kpc from both the nucleus of the S0 galaxy and the satellite galaxy, because we want to estimate the mass of the diffuse stellar and gaseous halo, which forms as an effect of the tidal encounter (a distance of 50 kpc is a conservative limit, to exclude matter that is still in the disc of the S0 galaxy or that can be re-accreted by the satellite galaxy). 

From Fig.~\ref{fig:figextra} it is apparent that $M_{\rm \ast,\,{}out}$ and $M_{\rm g,\,{}out}$ grow very fast after the first periapsis passage (the only exception being run~E). After the first Gyr, $M_{\rm \ast,\,{}out}$ and $M_{\rm g,\,{}out}$ remain almost constant, or keep growing slowly, till the end of the simulation. Thus, the first periapsis passage plays a major role in the formation of the stellar and gaseous halo. Furthermore, the halo is extremely long-lived. At the end of the simulation (10 Gyr after the first periapsis passage), $\sim{}70-95$\%   of the stars initially bound to the satellite galaxy lie in the stellar halo of the S0 galaxy (including the case of run~E). Similarly, $\sim{}60-80$\%  of the gas initially bound to the satellite galaxy ends up in the gaseous halo of the S0 galaxy at $t=10$ Gyr. The only exception is run~E, where $M_{\rm g,\,{}out}$ is only $\sim{}30$\%  of the initial gas mass at the end of the simulation.

\subsection{The stellar shells}

The main feature of the minor mergers in the stellar component of the S0 galaxy is the 
formation of shells (Figure~\ref{fig:fig6}). Shells are ubiquitous in the stellar component of our simulations: they start forming already during the first 
apoapsis passage, and last for the entire simulation in all runs. Our finding
 is consistent with the results of previous studies (e.g. \citealt{Peirani2010}).

\begin{figure}
\center{{
\epsfig{figure=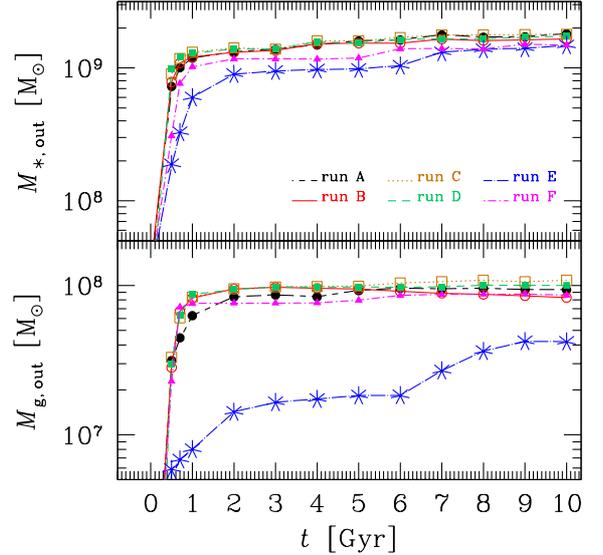,width=8.0cm} 
}}
\caption{\label{fig:figextra}
 Top (Bottom) panel: mass of stars (gas) initially bound to the satellite galaxy that lie(s) at a distance $>50$ kpc both from the centre of the S0 galaxy and from the centre of the satellite galaxy, as a function of time.  $M_{\rm \ast{},\,{}out}$ ($M_{\rm g,\,{}out}$)  can be considered as the mass of the stellar (gaseous) halo of the S0 galaxy. Time $t=0$ corresponds to the first periapsis passage. Long-short dashed black line and filled black circles: run~A; solid red line and open red circles: run~B; dotted ochre line and open ochre squares: run~C; dashed green  line and filled green squares: run~D; long-dash-dotted blue line and blue asterisks: run~E;  short-dash-dotted magenta line and filled magenta triangles: run~F. 
}
\end{figure}

\begin{figure}
\center{{
\epsfig{figure=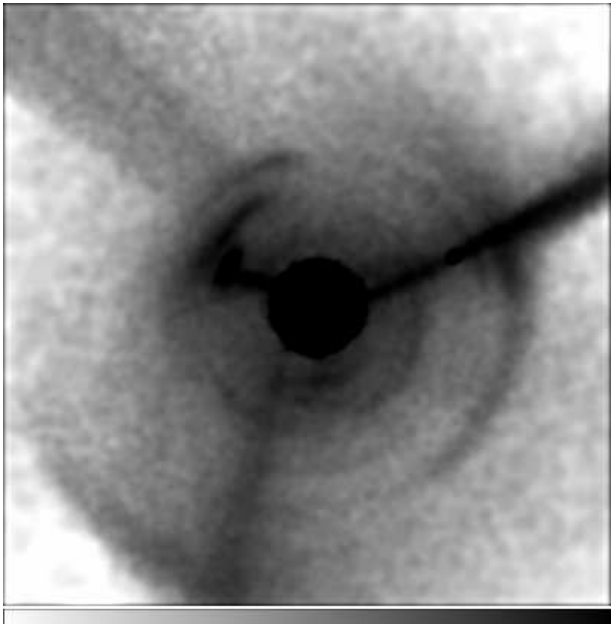,width=8.0cm} 
}}
\caption{\label{fig:fig6}
Projected density of  stars (logarithmic gray-scale map) 5 Gyr since the first periapsis passage in run~A. The box measures $400\times{}400$ kpc. The stellar disc of the S0 (seen face-on) is saturated.
}
\end{figure}

 In all simulations, the mass confined in the stellar shells is $\gtrsim{}0.5\,{}M_{\rm \ast{},\,{}out}$, i.e. it is more than half of the total mass of the stellar halo discussed in Section~\ref{sec:halo} (Fig.~\ref{fig:figextra}). On the other hand, the surface brightness of most shells is quite low. For example, the total mass of stars in the shells in run~A at $5$ Gyr is $\sim{}7\times{}10^8$ M$_\odot{}$, i.e. it is nearly 35\%{} of the initial stellar mass of the satellite galaxy, but this mass is spread over $\sim{}400\times{}400$ kpc. The maximum density in the strongest shells visible in Fig.~\ref{fig:fig6} is $\sim{}1.4-2.6\times{}10^4$ M$_\odot{}$ kpc$^{-2}$, much lower than the density in the outer parts of the S0 disc ($\sim{}10^7$ M$_\odot{}$ kpc$^{-2}$). 
At $7$ Gyr, the total mass of stars in the shells in run~A rises up to $\sim{}9\times{}10^8$ M$_\odot{}$, i.e. it is nearly 45\%{} of the initial stellar mass of the satellite galaxy, but the density in the single shells and the ratio between density of the stellar disc and density of the shells do not change.  Most shells in our simulations lie at much larger radii than the disc scale-length of the S0 galaxy, and are composed of matter removed from the satellite galaxy. We do not find evidence of shells composed of matter from the S0 galaxy (as proposed by \citealt{Wallin1988}).

\subsection{The star formation rate}
\begin{figure}
\center{{
\epsfig{figure=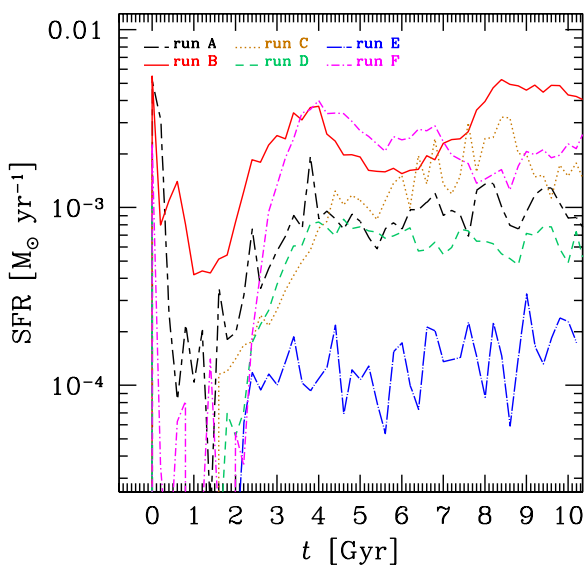,width=8.0cm} 
}}
\caption{\label{fig:fig7}
Star formation rate in all simulations, as a function of time. Time $t=0$ corresponds to the first periapsis passage. Long-short dashed black line: run~A; solid red line: run~B; dotted ochre line: run~C; dashed green  line: run~D; long-dash-dotted blue line: run~E; short-dash-dotted magenta line: run~F.
}
\end{figure}
Our simulations include recipes for stochastic SF, as described in Section~\ref{sec:method}. Because of our choice of not including gas in the initial conditions of the S0 galaxy, only the gas of the satellite galaxy can contribute to SF. This allows us to study SF from stripped gas in detail. Fig.~\ref{fig:fig7} shows the SF rate as a function of time in our simulations. In most simulations, SF starts at the first periapsis passage ($t=0$), and occurs mainly in the centre of the satellite galaxy, which has been strongly perturbed by the close interaction (see \citealt{Mapelli2012b,Mapelli2013a} for more details). As soon as gas is stripped and accretes onto the S0 galaxy, SF takes place mainly in the central parts of the S0 galaxy. Thus, the first SF burst  in Fig.~\ref{fig:fig7} occurs in the satellite galaxy during the first periapsis passage, while the second, longer SF episode occurs at the centre of the S0 galaxy, and is powered by stripped gas. We notice that the SF episode in the satellite galaxy and the SF episode in the centre of the S0 galaxy have a very different duration: they last for $<1$ Gyr and for $\sim{}8$ Gyr, respectively.

Interestingly, the SF rate is higher in runs~B, F (i.e. the prograde runs) and C. These are also the three runs where the ring is more short-lived (apart from run~E, in which the ring does not form). This suggests that the ring is suppressed when SF forms stars and stars explode as supernovae heating the surrounding gas. On the other hand, in our simulations most SF occurs in the nucleus of the S0 galaxy, rather than in the ring. This is likely an intrinsic limit of the stochastic recipes for SF (see e.g. \citealt{Mapelli2008,Mapelli2012b,Mapelli2013a}): these cannot resolve SF in relatively diffuse warm gas regions (the ring), while enhance SF in the density peaks (the nucleus). 

For most simulated galaxies, the SF rate is $\sim{}10^{-2}-10^{-3}$ M$_\odot{}$ yr$^{-1}$, almost constant for several Gyr. The run where gas stripping is less 
 efficient (run~E) is also the run with the lowest SF rate ($\sim{}10^{-4}$ M$_\odot$ yr$^{-1}$, 
 one order of magnitude lower than in the other runs).
Our simulated SF rate is quite lower than the SF rate of rejuvenated S0 galaxies as estimated by \citet{Salim2012} ($\sim{}0.5$ M$_\odot{}$ yr$^{-1}$). This difference can arise from the fact that our simulated S0 galaxies are completely gas-free before the merger, while real S0 galaxies have a gas reservoir, in which SF can be triggered during the merger.

\section{Discussion}

Gas rings form in most of our simulations. Thus, minor mergers with gas-rich satellites appear 
to be a very promising scenario for the formation of the gas rings observed in S0 galaxies. 
But how many of these S0 rings can be explained by minor mergers?

Cosmological simulations combined with semi-analytic models (e.g. \citealt{Bertone2009}) 
indicate that the minor-merger fraction at redshift $z=0$ (i.e. the fraction of galaxies 
that undergo a minor merger at $z=0$) is $f_{\rm mm}\sim{}0.1-0.3$,  for galaxies 
with stellar mass $M_{\ast{}}\sim{}0.3-1\times{}10^{11}$ M$_\odot{}$ (a reasonable 
mass range for S0 galaxies). 
Since the merger-induced SF episodes in our simulations are long-lived 
($\gtrsim{}6$ Gyr, Fig.~\ref{fig:fig7}), we expect that all S0 galaxies that 
undergo a minor merger at $z=0$ show signs of rejuvenation at present, 
under the reasonable assumption that all minor mergers occur with 
gas-rich satellites. Thus, we expect that $\sim{}10-30$\%   of S0 
galaxies appear somehow rejuvenated by a minor merger at $z=0$.

On the other hand, we must keep in mind several {\it caveats}  about this result. 
First, the minor-merger fraction estimated from cosmological simulations 
is quite uncertain. Furthermore, \citet{Bertone2009} show that the major-merger
 fraction estimated from the same cosmological simulations is in partial 
 disagreement with the data. The same issue might affect even the estimate of minor mergers.


We can compare such estimates based on cosmological simulations with several observational signatures of minor mergers.  
Shell structures  are one of the strongest scars leaved by merging episodes in ETGs. The literature provides quite different estimates about the fraction of  ETGs showing shells, since the shell phenomenon likely depends on the galaxy environment \citep[see e.g.][]{Marcum2004}. The fraction of ETGs showing systems of shells ranges from $\sim{}16$\%   (\citealt{Malin1983,Reduzzi1996}) up to $\sim{}56$\%  (according to \citealt{SeitzerSchweizer1990}, who study shell frequency among 74 field ETGs). A high fraction of  shells, 44\%  and 41\%, are found in ETGs by \citet{SchweizerFord1985} and by \citet{Colbert2001}, respectively. 

Recently, the ultra-deep survey {\tt MATLAS} investigated
92 out of 260 ETGs in the ATLAS$^{3D}$ sample \citep{Duc2014}. 
Shell structures are detected in 21\%  of the ETGs examined.  
Furthermore,  about half of the ETGs in \citet{Duc2014} sample show indications of minor merger (16\%), major merger (12\%), or interaction (22\%).  A large fraction of ETGs
show perturbed/asymmetric halos (35\%), streams (28\%) and tails (17\%).
In summary,  previous studies indicate that a significant  fraction of 
nearby ETGs, including S0 galaxies, may have suffered minor merger episodes.

A possible estimate of the fraction of ETGs that underwent a recent SF episode may be obtained from mid infrared (MIR) spectra.
 Polycyclic aromatic hydrocarbons (PAH) visible at MIR wavelengths are 
considered a clear signature of recent SF activity \citep{Tielens2008}. 
In ETGs, PAHs are detected  with both normal and anomalous
ratios \citep{Kaneda2008}.  \citet{Vega2010} suggest that 
anomalous PAH ratios are due to the injection of carbonaceous material
in the inter-stellar medium by carbon stars, which are present in the stellar populations with ages in the range of a few Gyr \citep{Nanni2013}. MIR spectra should then be able to provide indications about both on-going and
 recent (few Gyr) SF activity. 
Recently, \citet{Rampazzo2013} classified MIR Spitzer-IRS spectra in the 
nuclear region ($2-3$ $r_e/8$, where $r_e$ is the effective radius) of 99 ETGs 
according to galaxy activity. In general, elliptical galaxies are significantly more passive than S0 galaxies. 46$^{+11}_{-10}$\%  of elliptical galaxies and 20$^{+11}_{-7}$\%  of S0 galaxies show passively 
evolving spectra, i.e. spectra without any signature of either emission lines 
or  PAHs \citep[see also][]{Bressan2006}. PAHs 
are found in  47$^{+8}_{-7}$\%  of ETGs, but only 9$^{+4}_{-3}$\%  
have PAH ratios typical of star forming galaxies.   MIR spectra 
then suggest that a significant fraction of ETGs have been rejuvenated
in their {\it nuclear region} in the last few Gyr. Our simulations show that
both the ring and the nuclear region are interested by the rejuvenation
of the stellar population.

The aforementioned results are fairly consistent with the fact that about half of the S0 galaxies in the local Universe show signs of rejuvenation, 
and suggest that a significant fraction of these episodes in nearby S0 galaxies are induced by minor mergers with gas-rich satellites.

Furthermore, \citet{Annibali2007} modelled the nuclear Lick line-strength
 indices 65 nearby ETGs \citep{Rampazzo2005,Annibali2006}. 
 By comparing the number of rejuvenated ETGs with the total number of galaxies in the sample,  and by means of simple two-component single stellar population models, they estimate that the rejuvenation episodes do not involve more than 25\%  of the total galaxy mass.
This estimate, together with morphological signatures and MIR indication, witnesses that the role of minor mergers could be very important in the evolution of ETGs in general, and of S0s in particular.

 



How efficient is gas transfer in minor mergers? Based on the WHISP survey (The Westerbork HI survey of irregular and spiral galaxies, \citealt{vanderHulst2001}), \citet{Sancisi2008}  estimate that $\sim{}25$\%   of local galaxies show signs of possible minor mergers (which is in agreement with the aforementioned value from cosmological simulations), but find that the mass accretion rate is only $\sim{}1/10$ of the SF rate. 
Recently, \citet{DiTeodoro2014} estimate that 22\%  of galaxies in a sub-sample of the WHISP have a dwarf companion, but the maximum gas accretion rate from these companions is about five times lower than the average SF rate of the sample (see also \citealt{Holwerda2011}). These results suggest that minor mergers do not play a significant role in the total gas accretion budget. On the other hand, the galaxies considered by both \citet{Sancisi2008} and \citet{DiTeodoro2014} are gas-rich spiral galaxies. 

Spiral galaxies have their own reservoir of gas, and SF episodes might be triggered in several ways. 
In contrast, S0 galaxies do not have a rich reservoir of gas of their own: for S0 galaxies, the infall 
of a gas-rich companion might be the privileged way to refuel and re-activate SF.



\section{Conclusions}
In this paper, we investigated interactions between a S0 galaxy and a gas-rich small satellite galaxy, 
by means of N-body/SPH simulations. The satellite galaxy is initially on a nearly parabolic orbit with
 respect to the primary galaxy. We ran six simulations, varying the impact parameter (10 and 30 kpc), 
 the inclination with respect to the plane of the S0 disc (0, 45 and 90$^\circ{}$), and the orientation
  of the orbital angular momentum of the secondary galaxy with respect to the spin of the primary
   galaxy (prograde and retrograde).  In all simulations, the satellite galaxy is slowly disrupted by the 
   interaction, its gas is stripped during each periapsis passage, and accretes onto the S0 galaxy.

We find that about $1/10$ of the total gas mass of the satellite galaxy ends up in the 
central region of the S0 galaxy. Warm dense gas rings form in most our simulations, 
indicating that minor mergers are a viable scenario for the formation of gas rings in 
S0 galaxies and for the rejuvenation of S0 galaxies. The radius of the ring is of $\sim{}6-13$ kpc, 
depending on the simulation, and the total mass confined in the ring is $1/100-1/10$ 
of the initial gas mass of the satellite galaxy.

We find that gas is stripped earlier in prograde runs than in retrograde ones, because 
tidal stripping is more efficient in prograde runs, where the relative velocity between 
the satellite galaxy and the particles in the S0 disc is lower. Thus, prograde encounters build 
gas rings earlier, but these rings have a shorter life ($1-3$ Gyr) than those formed in 
either retrograde encounters or non-coplanar encounters. Rings formed in  retrograde
 encounters and non-coplanar encounters can live unperturbed for more than $\sim{}6$ Gyr. 

Furthermore, the rings keep memory of the orbit of the satellite galaxy: rings formed in 
prograde encounters are co-rotating with the disc of the S0 galaxy, while rings formed in retrograde 
encounters are counter-rotating. Similarly, satellite galaxies whose orbit was coplanar
 with respect to the disc of the S0 galaxy form co-planar rings, while satellite galaxies 
 whose orbit was inclined by 45 and 90$^\circ{}$ with respect to the disc of the primary 
 galaxy form rings that are inclined by 45 and 90$^\circ{}$, respectively.

This result has another crucial implication: rings form and grow even without bar resonances. 
In our simulations, the strongest and most long-lived ring is polar (run~D): it forms with an 
inclination angle of 90$^\circ$.

Furthermore, the impact parameter appears to be a very important ingredient: if it is too 
large (30 kpc) and if the run is retrograde, the stripping is very inefficient and rings do not
 form for the entire duration of the simulation ($>10$ Gyr).

The formation of stellar shells surrounding the S0 galaxy is also ubiquitous in our simulations, 
but such shells might be very faint: the mass density of stars in the shells is about three orders 
of magnitude lower than that in the disc of the S0 galaxy.

During the first periapsis passage, SF occurs mainly in the disc of the satellite galaxy, 
which is strongly perturbed by the interaction. As soon as the stripped gas reaches the 
centre of the S0 galaxy, SF switches on even in the S0 galaxy, while it is fast quenched 
in the satellite galaxy, whose initial gas was in large portion removed. The stripped gas 
powers an extremely long-lived ($\sim{}8$ Gyr) episode of SF in the central region of the 
S0 galaxy. The SF rate is $\sim{}10^{-4}-0.01$ M$_\odot$ yr$^{-1}$, depending on the 
simulation. We expect that it can reach higher values for a larger initial gas mass of the 
satellite galaxy and/or if we account for the gas reservoir of the S0 galaxy. SF and stellar feedback (supernova explosions) might play a strong 
role in the suppression of the gas rings: we find that gas rings disappear earlier in the 
runs where SF starts earlier and  reaches the highest level (runs~B and F).


Finally, we discuss our results in view of the minor-merger fraction derived from cosmological simulations \citep{Bertone2009}, from
morphological signatures (shells and streams) investigated in deep surveys \citep{Duc2014},
and from the presence of PAHs in the nuclear region of S0 galaxies.
We find that minor mergers might account for a significant fraction of rejuvenated S0 galaxies.
 This result is a fundamental clue for our knowledge of rejuvenation processes in ETGs, but 
deserves further investigations, since it relies on a very uncertain estimate of the minor-merger
 rate in the local Universe, and on a limited sample of galaxies investigated with  MIR and  optical spectroscopy. 

%
%
%
%
%
%
%

\section*{Acknowledgments}
We thank the referee, Curtis Struck, for his helpful comments. We also thank the authors of gasoline, especially J. Wadsley, T. Quinn and J. Stadel. We thank L.~Widrow for providing us the code to generate the initial conditions, and A. Moiseev, E. Iodice, M. Spavone, E. Ripamonti and L. Mayer for useful discussions. To analyze simulation outputs, we made use of the software TIPSY\footnote{\tt http://www-hpcc.astro.washington.edu/tools/tipsy/\\tipsy.html}.
The simulations were performed with the {\it lagrange} cluster at CILEA and with the PLX, Eurora and Fermi cluster at CINECA. We acknowledge the CINECA Award N. HP10CLI3BX, HP10B3BJEW and HP10B338N6 for the availability of high performance computing resources and support. MM acknowledges financial support from the Italian Ministry of Education, University and Research (MIUR) through grant FIRB 2012 RBFR12PM1F, and from INAF through grant PRIN-2011-1.


\end{document}